\documentclass[preprint,1p]{elsarticle}

\usepackage{hyperref}

\journal{XXXXX}

\usepackage {amsmath}
\usepackage{amssymb}

\newproof{pf}{Proof}
\newdefinition{rmk}{Remark}
\usepackage{graphicx}
\usepackage{subfigure}
\usepackage{color}

\begin{document}

\begin{frontmatter}

\title{Entangled state generation via quantum walks with multiple coins}

\author[a,b]{Meng Li}

\author[a,c]{Yun Shang\corref{cor1}}
\ead{shangyun602@163.com}

\cortext[cor1]{Corresponding author}

\address[a]{Institute of Mathematics, Academy of Mathematics and Systems Science, Chinese Academy of Sciences, Beijing 100190, China}
\address[b]{School of Mathematical Sciences, University of Chinese Academy of Sciences, Beijing 100049, China}
\address[c]{NCMIS, MDIS, Academy of Mathematics and Systems Science, Chinese Academy of Sciences, Beijing 100190, China}

\begin{abstract}
Generation of entangled state is of paramount importance both from  quantum theoretical foundation and technology applications. Entanglement swapping provides an efficient method to generate entanglement in quantum communication protocols.
However, perfect Bell measurements for qudits, the key to entanglement swapping, have been proven impossible to achieve by using only linear elements and particle detectors.
To avoid this bottleneck, we propose a novel scheme to generate entangled state including two-qubit entangled state, two-qudit entangled state, three-qubit GHZ state and three-qudit GHZ state between several designate parties via the model of quantum walks with multiple coins.
 Then we conduct experimental realization of Bell state and three-qubit GHZ state between several designate parties on IBM quantum platform and the result has high fidelity by preforming quantum tomography.
In the end, we give a practical application of our scheme in multiparty quantum secret sharing.
\end{abstract}


\end{frontmatter}

\section{Introduction}
Entanglement, a very powerful and efficient quantum resource, is a cornerstone of many quantum communication and quantum computation protocols.
A lot of quantum schemes depend heavily on the property of entanglement \cite{horodecki2009entanglement}, such as quantum key distribution \cite{ekert1991qkd}, quantum teleportation \cite{bennett1993teleporting}, quantum metrology \cite{giovannetti2006metrology}, and quantum sensing \cite{degen2017quantumsensing}.

There is no doubt that the preparation of entangled quantum states is of great concern in the past decades.
Many schemes have been proposed theoretically and achieved experimentally \cite{dicarlo2010preparation, neeley2010generation, delteil2016generation, Hu2020entangledqudits, Erhard2020review}.
However, the preparation of high-dimensional entangled states is so difficult that it has been implemented almost exclusively in photonic systems \cite{krenn2014generation}.
Entanglement swapping \cite{zukowski1993entanglementswapping}, also known as teleportation of entanglement, can entangle two particles that are not related at first by performing a joint Bell-state measurement on them, which plays an important role in long distance quantum communication, such as quantum repeater \cite{briegel1998repeater, santra2019repeater}. It is also the heart for quantum network and entanglement distribution.
However, Ref. \cite{John2002generalizedmeasurment} has pointed out that it is impossible to perform perfect Bell state measurement for qudits by using only linear elements and particle detectors.
It is of interest to know if we can sidestep this thorny issue but still entangle designate parts.
And in addition to the simple case involving two-qubit entangled quantum state, GHZ entanglement swapping by using three pairs EPR entangled state has also been discussed \cite{Lu2009GHZ}. Then, Ref. \cite{Su2016multipartite} demonstrated entanglement swapping between two multipartite entangled state, such as GHZ state and EPR state, and two GHZ states.
However, a relatively general framework for generating entangled state via entanglement swapping is missing, let alone the experimental platform chosen to test.

Motivated by this, we develop a new scheme to generate multipartite entanglement in high dimensions by using quantum walks instead of direct Bell state measurement in the framework of entanglement swapping and thus avoid the difficulty of the realization of Bell state measurement.
Our scheme is mainly based on the model of quantum walks with multiple coins \cite{brun2003quantum} which can bring out entanglement between position space and coin space and has been proved to be useful in quantum communication protocols\cite{wang2017generalized, shang2019pst}.
Here we mainly consider quantum walks with multiple coins on 2-line, 2-circle and 2-complete ($d$-complete) graph, which are all the building blocks in quantum network.
In this way, we can generate two-qubit entangled  state, two-qudit entangled state, three-qubit GHZ state and even three-qudit GHZ state between several designate parties.

In terms of experimental realization of entangled state, it is very difficult to find an appropriate experimental preparation device especially for the high-dimensional entangled state, let alone universal method and experimental device.
However, the physical implementations of the quantum walk have been realized in many different physical systems, such as trapped atom \cite{karski2009trappedatom}, trapped ions \cite{tamura2020trappedions}, photonic system \cite{tang2018photonic} and so on. Since quantum walk is a universal quantum computing model \cite{childs2009universal, childs2013universal, lovett2010universal, underwood2010universal}, these schemes may provide a universal platform for entanglement generation.

Recently, IBM Quantum Experience, as a superconducting flux qubit based open platform, attracts the attention of many researchers \cite{IBM_Q}. Many theoretical schemes in the field of quantum computation and quantum information have been detected and performed on this platform \cite{shang2019experiment, alsina2016experimental, devitt2016experiment, behera2017experimental, Wang2018entangledqubits}.
Here, we carry out experiments on this platform to generate Bell state and three-qubit GHZ state between several designate parties.
Also, the accuracy of our scheme can be verified well by performing quantum state tomography \cite{james2001measurement}.

Furthermore, we give its application in quantum communication. In detail, we provide an improved multiparty quantum secret sharing protocol based on Ref. \cite{zhang2005MQSS} by using our scheme.
Compared with the original protocols, our new scheme are high code capacity and high security. And it is easy to be implemented in experiment.

\section{Results}
\begin{figure}[htb]
 \centering
 \subfigure[]{\label{schematic_diagram1}
 \includegraphics[width=5cm]{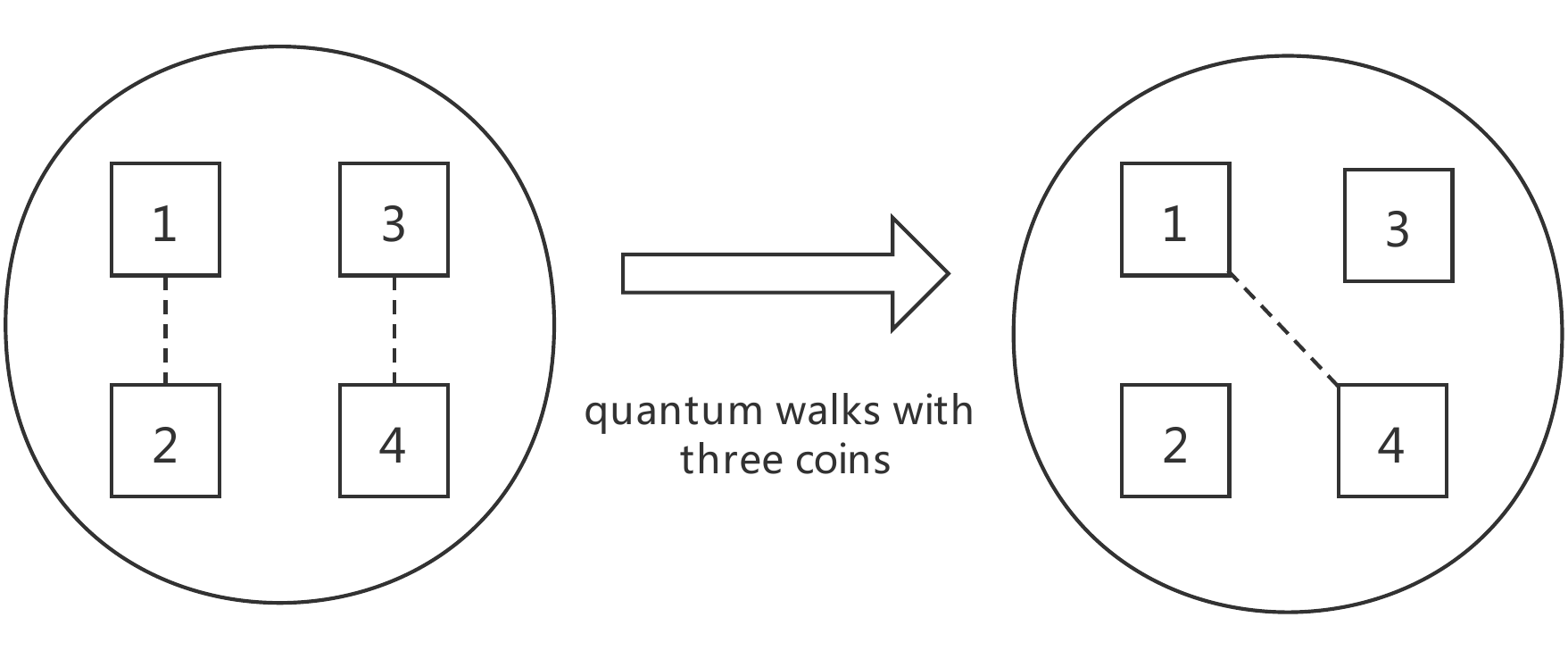}}\ \ \
 \subfigure[]{\label{schematic_diagram2}
 \includegraphics[width=5.5cm]{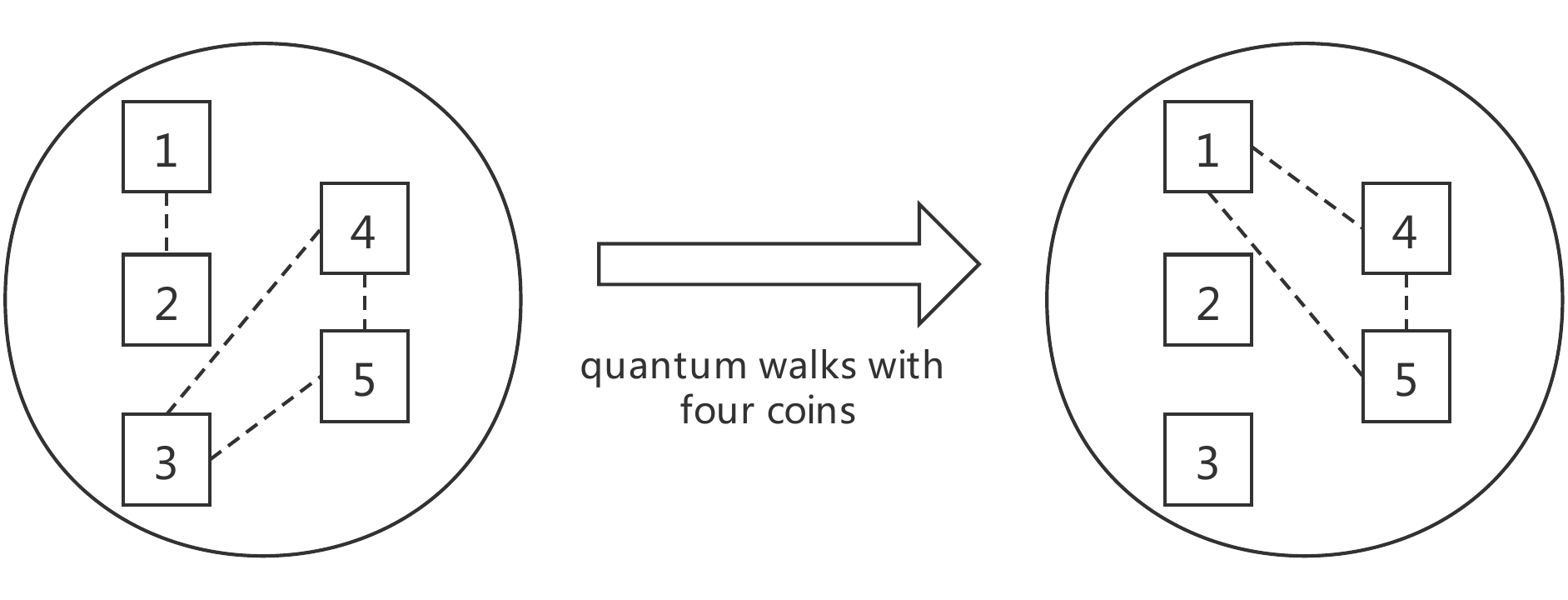}}
 \caption{The schematic diagram of our scheme. The small boxes indicate particles and the dotted lines indicate entanglement. (a) represents the generation of bipartite entangled state. (b) represents the generation of GHZ state.
 Due to the symmetry of the particles, any two or three unrelated particles can be entangled by our scheme.
But for the sake of narration, let us take entanglement generation of particles $1, 4$ and particles $1, 4, 5$ as examples.}
 \label{diagram_scheme}
 \end{figure}

For the quantum walks with $k$ coins on an arbitrary graph, the $k$ coins were used cyclically.
At $i$-th step of quantum walks, the coin state can be flipped with the $i$-th coin operator $C_{i}$ and then the walker moves to another position according to the conditional shift operator $S_{i}$, which can be described as
\begin{equation}
U_{i}=(S_{i}\otimes I)\cdot(I\otimes C_{i}).
\end{equation}
After $t$ steps, the initial state $|\phi(0)\rangle$ evolves into
\begin{equation}
|\phi(t)\rangle=(U_{k}\cdots U_{1})^{t/k}|\phi(0)\rangle.
\end{equation}
We denote the line, circle and complete graph with $N$ vertices as $N$-line, $N$-circle and $N$-complete graph respectively. The three corresponding conditional shift operators acting on the position space and the coin space that is active at the step can be written as:
\begin{equation} \label{Ncomplete}
S^{N-complete}=\sum_{x,j=0}^{N-1}|(x+j)\bmod\,N\rangle\langle x|\otimes|j\rangle\langle j|,
\end{equation}
\begin{equation} \label{Ncycle}
S^{N-circle}=\sum_{x=0}^{N-1}(|(x+1)\bmod\,N\rangle\langle x|\otimes|0\rangle\langle0|+|(x-1)\bmod\,N\rangle\langle x|\otimes|1\rangle\langle1|),
\end{equation}
\begin{equation} \label{Nline}
\begin{split}
S^{N-line}&=\sum_{x=0}^{N-2}|(x+1)\bmod\,N\rangle\langle x|\otimes|0\rangle\langle0|+\sum_{x=1}^{N-1}|(x-1)\bmod\,N\rangle\langle x|\otimes|1\rangle\langle1| \\
&+|N-1\rangle\langle N-1|\otimes|1\rangle\langle0|+|0\rangle\langle0|\otimes|0\rangle\langle1|.
\end{split}
\end{equation}
In particular, when the number of vertices is $2$, the conditional shift operators are:
\begin{equation} \label{2complete}
S^{2-complete}=(|0\rangle\langle0|+|1\rangle\langle1|)\otimes|0\rangle\langle0|
              +(|1\rangle\langle0|+|0\rangle\langle1|)\otimes|1\rangle\langle1|,
\end{equation}
\begin{equation} \label{2cycle}
S^{2-circle}=(|1\rangle\langle0|+|0\rangle\langle1|)\otimes|0\rangle\langle0|
           +(|1\rangle\langle0|+|0\rangle\langle1|)\otimes|1\rangle\langle1|,
\end{equation}
\begin{equation} \label{2line}
S^{2-line}=|1\rangle\langle0|\otimes|0\rangle\langle0|+|0\rangle\langle1|\otimes|1\rangle\langle1|
          +|0\rangle\langle0|\otimes|0\rangle\langle1|+|1\rangle\langle1|\otimes|1\rangle\langle0|.
\end{equation}

\subsection{Generation of two-qubit entangled state}
We start by discussing the generation of two-qubit entangled state. And the experimental setup is illustrated in Figure \ref{schematic_diagram1}.
Without loss of generality, we can assume that the initial state of the four particles is
\begin{equation} \label{qubit_initial}
|\psi(0)\rangle=(a|01\rangle+b|10\rangle)_{1,2}(a|01\rangle+b|10\rangle)_{3,4},
\end{equation}
where $|a|^{2}+|b|^{2}=1$.
Now we try to generate entanglement between particle $1$ and $4$.
Here, we can view particle $1$ and $2, 3, 4$ as walker and three coins respectively.
And thus we can consider quantum walks with three coins on $2$-line, $2$-circle and $2$-complete graph.

For the case of quantum walks with three coins on $2$-line, we can perform two-step quantum walks where $C_{1}=H$ and $C_{2}=I$. According to conditional shift operator and initial state shown in equation (\ref{2line}) and (\ref{qubit_initial}), the evolution process is written as follows:
\begin{equation}
\begin{split}
|\psi(1)\rangle=&\frac{1}{\sqrt{2}}(a^{2}|1001\rangle-a^{2}|0001\rangle+ab|1010\rangle-ab|0010\rangle\\
&+ab|1101\rangle+ab|0101\rangle+b^{2}|1110\rangle+b^{2}|0110\rangle)_{1,2,3,4},
\end{split}
\end{equation}
\begin{equation}
|\psi(2)\rangle=\frac{1}{\sqrt{2}}(-a|00\rangle+a|01\rangle+b|10\rangle+b|11\rangle)_{2,3}(a|11\rangle+b|00\rangle)_{1,4}.
\end{equation}
Therefore, we can definitely obtain entangled state between particle $1$ and $4$ by performing two local measurements on the other two particles. In particular, the state will be Bell state, also known as maximal entangled state, when $a=b=\frac{1}{\sqrt{2}}$.

In the case of quantum walks with three coins on $2$-circle, it is impossible to get entangled between particle $1$ and $4$ no matter how many steps we take. In this case, the conditional shift operator happens to be the second-order Pauli matrix $X_{2}$.
Therefore, the general form of evolved quantum state under three-step quantum walks is $(I\otimes C_{1}\otimes C_{2}\otimes C_{3})(a^{2}|1101\rangle+ab|1110\rangle+ab|0001\rangle+b^{2}|0010\rangle)$.
And it has a definite explicit form by giving three unitary coin operators with parameters.
Then, it is not hard for us to find that there is no satisfied coin operator by analyzing the final state with parameters.

As regards quantum walks with three coins on $2$-complete graph, we can take three-step quantum walks where $C_{1}=I$, $C_{2}=H$ and $C_{3}=X_{2}$. Thus
the whole evolutionary process can be described as follows:
\begin{equation}
|\psi(1)\rangle=(a^{2}|1101\rangle+ab|1110\rangle+ab|1001\rangle+b^{2}|1010\rangle)_{1,2,3,4},
\end{equation}
\begin{equation}
\begin{split}
|\psi(2)\rangle=&\frac{1}{\sqrt{2}}(a^{2}|1101\rangle+a^{2}|0111\rangle+ab|1100\rangle-ab|0110\rangle\\
&+ab|1001\rangle+ab|0011\rangle+b^{2}|1000\rangle-b^{2}|0010\rangle)_{1,2,3,4},
\end{split}
\end{equation}
\begin{equation} \label{final_state_of_qubit_on_complete}
|\psi(3)\rangle=\frac{1}{\sqrt{2}}[(b|00\rangle+a|10\rangle)_{2,3}(a|10\rangle+b|01\rangle)_{1,4}+(b|01\rangle+a|11\rangle)_{2,3}(a|00\rangle-b|11\rangle)_{1,4}].
\end{equation}
So we can obtain entanglement between particle $1$ and $4$ by performing three-step quantum walks and two local measurements.
Of course, the state obtained happens to be Bell state when $a=b=\frac{1}{\sqrt{2}}$.
In addition, there are other ways to select coin operators to generate new entanglement. For example, $C_{1}=I$, $C_{2}=X_{2}$, $C_{3}=H$ or $C_{1}=X_{2}$, $C_{2}=X_{2}$, $C_{3}=H$, or $C_{1}=X_{2}$, $C_{2}=H$, $C_{3}=X_{2}$ can also be used to generate entanglement between particle $1$ and $4$.

\subsection{Generation of two-qudit entangled state}
High-dimensional entangled states are of paramount interest both from a theoretical and practical perspective.
Generalized Bell states in Hilbert space $C^{d}\otimes C^{d}$, a basis of maximally entangled high-dimensional bipartite states, can be described as \cite{bennett1993teleporting, sych2009generalizedBellstates}
\begin{equation}
|\psi_{k,l}\rangle=\frac{1}{\sqrt{d}}\sum_{m=0}^{d-1}\exp(\frac{2\pi \mathrm{i}}{d}mk)|m\rangle|m-l\rangle,
\end{equation}
where $m-l$ means $(m-l)\ \mathrm{mod}\ d$.
The general entangled state can be written as $\sum_{i=0}^{d-1}a_{i}|i\rangle|i\rangle$, where $\sum_{i=0}^{d-1}|a_{i}|^{2}=1$.
For the sake of simplicity, we denote $d$-order generalized Pauli operator as $X_{d}=\sum_{i=0}^{d-1}|i+1\rangle\langle i|$, which is also known as discrete Weyl operator.
Next, we propose two novel schemes to generate maximal and non-maximal entangled high-dimensional states (two-qudit entangled state) by performing quantum walks with three coins on $d$-complete graph.  And its schematic diagram is shown in Figure \ref{schematic_diagram1}.

For the case of maximal entangled state, the initial state is $|\psi(0)\rangle=|\psi_{k,l}\rangle_{1,2}|\psi_{k,l}\rangle_{3,4}=\frac{1}{d}\sum_{m,n=0}^{d-1}\exp(\frac{2\pi\mathrm{i}}{d}(m+n)k)|m,m-l,n,n-l\rangle$.
Now we can perform three-step quantum walks where $C_{1}=I$, $C_{2}=F$ (Fourier transformaiton) and $C_{3}=X_{d}$.
Thus, we can obtain the final state:
\begin{equation}
|\psi(3)\rangle=\frac{1}{d}\sum_{m,p=0}^{d-1}[\exp(\frac{2\pi\mathrm{i}}{d}mk)|m-l\rangle_{2}|p\rangle_{3}
(\frac{1}{\sqrt{d}}\sum_{n=0}^{d-1}\exp(\frac{2\pi\mathrm{i}}{d}(k+p)n)|2m+p+n-2l+1\rangle_{1}|n-l+1\rangle_{4})].
\end{equation}
Then we take two local measurements on particle $2$ and $3$ and assume that the result are $\exp(\frac{2\pi\mathrm{i}}{d}m_{0}k)|m_{0}-l\rangle_{2}|p_{0}\rangle_{3}$. So the reduced state would be $\exp(-\frac{2\pi\mathrm{i}}{d}(2m_{0}+p_{0}-2l+1)(k+p_{0}))|\psi_{k+p_{0},2m_{0}+p_{0}-l}\rangle_{1,4}$.

More generally, in the case of non-maximal entangled state, the initial state can be written as $|\psi(0)\rangle=(\sum_{i=0}^{d-1}a_{i}|i\rangle|i\rangle)_{1,2}\otimes(\sum_{j=0}^{d-1}b_{j}|j\rangle|j\rangle)_{3,4}$, where $\sum_{i=0}^{d-1}|a_{i}|^{2}=\sum_{j=0}^{d-1}|b_{j}|^{2}=1$.
After the same three-step quantum walks ($C_{1}=I$, $C_{2}=F$ and $C_{3}=X_{d}$), the final state will be
$|\psi(3)\rangle=\frac{1}{\sqrt{d}}\sum_{i,k=0}^{d-1}[a_{i}|i\rangle_{2}|k\rangle_{3}(\sum_{j=0}^{d-1}b_{j}\exp(\frac{2\pi\mathrm{i}}{d}jk)|2i+j+k+1\rangle_{1}|j+1\rangle_{4})]$.
Similarly, particle $1$ is entangled with $4$ after two local measurements.

\subsection{Generation of three-qubit GHZ state}
The three-qubit GHZ states  maximize entanglement monotones \cite{normalform2003, threequbit2000} and have extremely non-classical properties.
The setup for this problem is shown in the Figure \ref{schematic_diagram2}.
The initial state of the five particles can be written as
\begin{equation} \label{GHZ_initial}
|\psi(0)\rangle=(a|01\rangle+b|10\rangle)_{1,2}(\frac{|000\rangle+|111\rangle)}{\sqrt{2}})_{3,4,5},
\end{equation}
where $|a|^{2}+|b|^{2}=1$.
In order to generate entanglement between particle $1$, $4$ and $5$, we can view these five particles as the walker and four coins respectively.
So we can discuss the quantum walks with four coins on $2$-line, $2$-circle and $2$-complete graph.

Let us start by talking about the case of quantum walks with four coins on $2$-line. And in order to do that, we perform two-step quantum walks where $C_{1}=H$ and $C_{2}=X$. Thus the initial quantum state presented in equation (\ref{GHZ_initial}) will evolves over time into
\begin{equation}
|\psi(1)\rangle=\frac{1}{2}(-a|00\rangle+b|01\rangle+a|10\rangle+b|11\rangle)_{1,2}(|000\rangle+|111\rangle)_{3,4,5},
\end{equation}
and
\begin{equation}
|\psi(2)\rangle=\frac{1}{2}(-a|00\rangle+a|01\rangle+b|10\rangle+b|11\rangle)_{2,3}(|000\rangle+|111\rangle)_{1,4,5}.
\end{equation}
Then we just have to make local measurements of particle $2$ and particle $3$, and no matter what the measurements are, the remaining three particles always collapses to GHZ state.

For quantum walks with four coins on $2$-circle, particle $1$, $4$ and $5$ will never get entangled no matter how many steps we take.
Given an initial state in equation (\ref{GHZ_initial}), after four steps the state evolves to $\frac{1}{\sqrt{2}}(I\otimes C_{1}\otimes C_{2}\otimes C_{3}\otimes C_{4})(a|01000\rangle+a|01111\rangle+b|10000\rangle+b|10111\rangle)$, where these four operators can be any unitary operator of the second order.
By calculating the undetermined coefficients of these four coin operators, we find that particle $1$, $4$ and $5$ are never completely entangled.

In regard to the quantum walks with four coins on $2$-complete graph, we can generate entanglement between particle 1, 4 and 5 by performing four-step quantum walks where $C_{1}=C_{2}=I$ and $C_{3}=C_{4}=H$.
After the first two steps of quantum walks, the quantum state evolves into
\begin{equation}
|\psi(2)\rangle=\frac{1}{\sqrt{2}}(a|11000\rangle+a|01111\rangle+b|10000\rangle+b|00111\rangle)_{1,2,3,4,5}.
\end{equation}
The resulting state will be
\begin{equation}
\begin{split}
|\psi(3)\rangle=&\frac{1}{2}(a|11000\rangle+a|01010\rangle+a|01101\rangle-a|11111\rangle\\
&+b|10000\rangle+b|00010\rangle+b|00101\rangle-b|10111\rangle)_{1,2,3,4,5}
\end{split}
\end{equation}
after one step with coin operator $C_{3}=H$.
And then the final quantum state can be described as follows:
\begin{equation}
\begin{split}
|\psi(4)\rangle=\frac{1}{2\sqrt{2}}[&(b|00\rangle+a|10\rangle)_{2,3}(|100\rangle+|001\rangle+|010\rangle+|111\rangle)_{1,4,5}\\
&+(b|01\rangle+a|11\rangle)_{2,3}(|000\rangle-|101\rangle-|110\rangle+|011\rangle)_{1,4,5}].
\end{split}
\end{equation}
So we can get the GHZ-like state \cite{yang2009GHZlike} of particle $1$, $4$ and $5$, which is the entangled quantum state and belongs to the class of GHZ state, by taking two local measurements of particle $2$ and $3$.

\subsection{Generation of three-qudit GHZ state}
Three-qudit GHZ state can be written as $\frac{1}{\sqrt{d}}\sum_{n=0}^{d-1}|n\rangle^{\otimes^{3}}$.
And its generation process is presented in Figure \ref{schematic_diagram2}.
For the initial state
\begin{equation}
|\psi(0)\rangle=
|\psi_{k,l}\rangle_{1,2}(\frac{1}{\sqrt{d}}\sum_{n=0}^{d-1}|n\rangle^{\otimes^{3}})_{3,4,5}
=\frac{1}{d}\sum_{m,n=0}^{d-1}\exp(\frac{2\pi\mathrm{i}}{d}mk)|m,m-l,n,n,n\rangle_{1,2,3,4,5},
\end{equation}
we can perform four-step quantum walks with four coins on $d$-complete graph by choosing $C_{1}=C_{2}=I$ and $C_{3}=C_{4}=F$ to establish the entanglement between particle 1, 4 and 5 that are unrelated at first.
The evolutionary process can be described as follows:
\begin{equation}
|\psi(2)\rangle=\frac{1}{d}\sum_{m,n=0}^{d-1}\exp(\frac{2\pi\mathrm{i}}{d}mk)|2m-l+n,m-l,n,n,n\rangle_{1,2,3,4,5},
\end{equation}
\begin{equation}
\begin{split}
|\psi(4)\rangle&=\frac{1}{d^{2}}\sum_{m,n,p,q=0}^{d-1}\exp(\frac{2\pi\mathrm{i}}{d}(mk+np+nq))|2m-l+n+p+q,m-l,n,p,q\rangle_{1,2,3,4,5}\\
&=\frac{1}{d}\sum_{m,n=0}^{d-1}[\exp(\frac{2\pi\mathrm{i}}{d}mk)|m-l\rangle_{2}|n\rangle_{3}
(\frac{1}{d}\sum_{p,q=0}^{d-1}\exp(\frac{2\pi\mathrm{i}}{d}(np+nq))|2m-l+n+p+q,p,q\rangle_{1,4,5})].
\end{split}
\end{equation}
Then we can obtain an entangled state of particle $1$, $4$ and $5$ by performing two local measurements on particle $2$ and $3$.
Note that any state like $\sum_{p,q=0}^{d-1}|x_{0}+p+q,p,q\rangle$ is completely entangled, where $x_{0}\in\{0,1,2,\cdots,d-1\}$. It can be seen as the high-dimensional case of GHZ-like state.

\subsection{Experimental protocol}
We provide the experimental realization of our scheme using IBM quantum computer $ibmq_{\_}5_{\_}yorktown-ibmqx2$ (it is called $ibmqx2$ for short) and $simulator$ with 8192 number shots (see Supplementary Note 1 for the introduction). Here, we mainly discuss the generation of standard Bell state and three-qubit GHZ state based on quantum walks.

\subsubsection{Experimental generation of Bell state on $2$-complete graph}
\begin{figure}[htb]
 \centering
 \includegraphics[width=9cm]{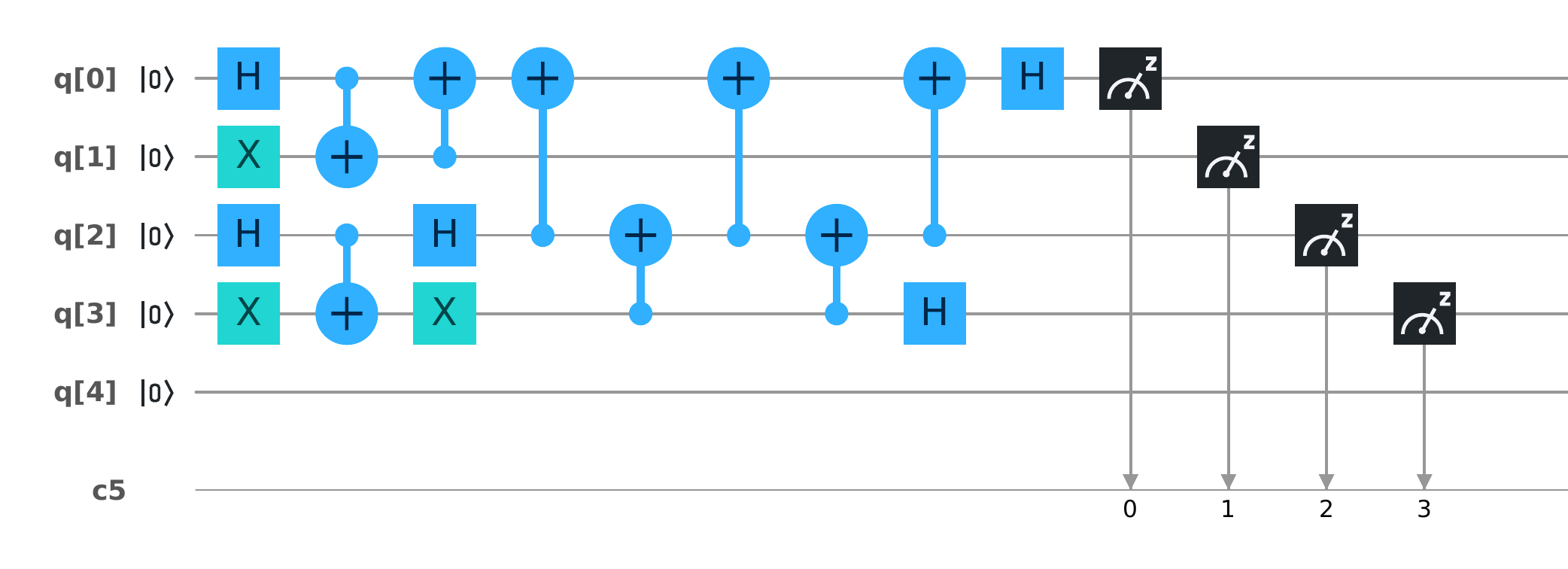}
 \caption{Quantum circuit for generating Bell state between two parties that are not related at first. The measurements on $q[0]$ and $q[3]$ are in X-basis. The measurements on $q[1]$ and $q[2]$ are in Z-basis.}
 \label{ES_Bell_circuit}
 \end{figure}
\begin{figure}[htb]
 \centering
 \subfigure[]{\label{ES_Bell_result_simulator}
 \includegraphics[width=6cm]{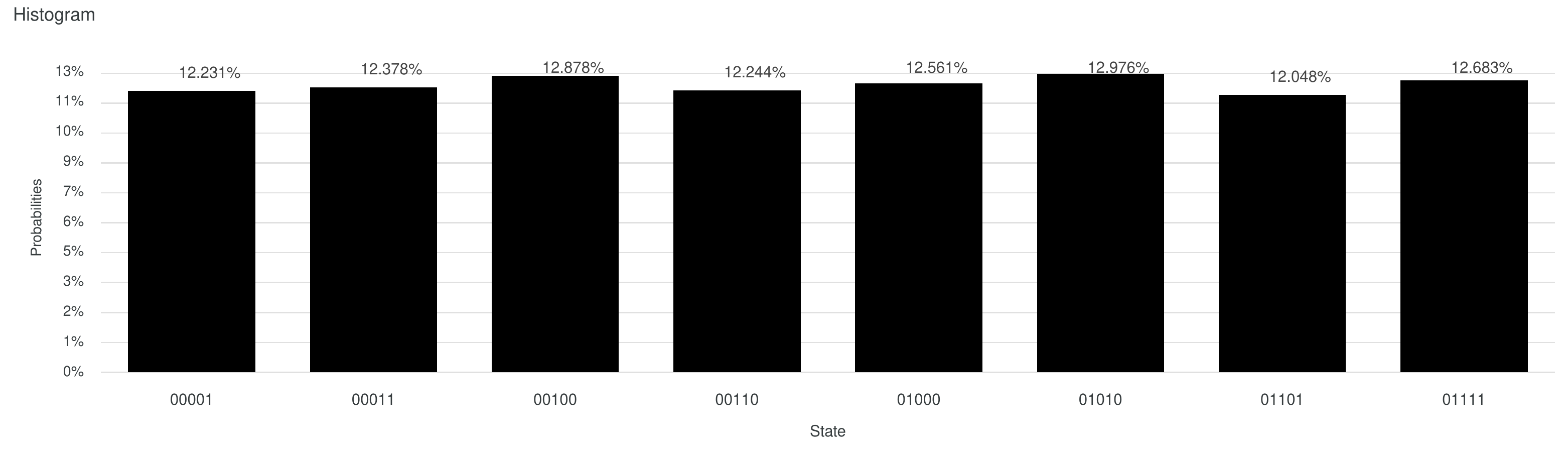}}\ \
 \subfigure[]{\label{ES_Bell_result_ibmqx2}
 \includegraphics[width=6cm]{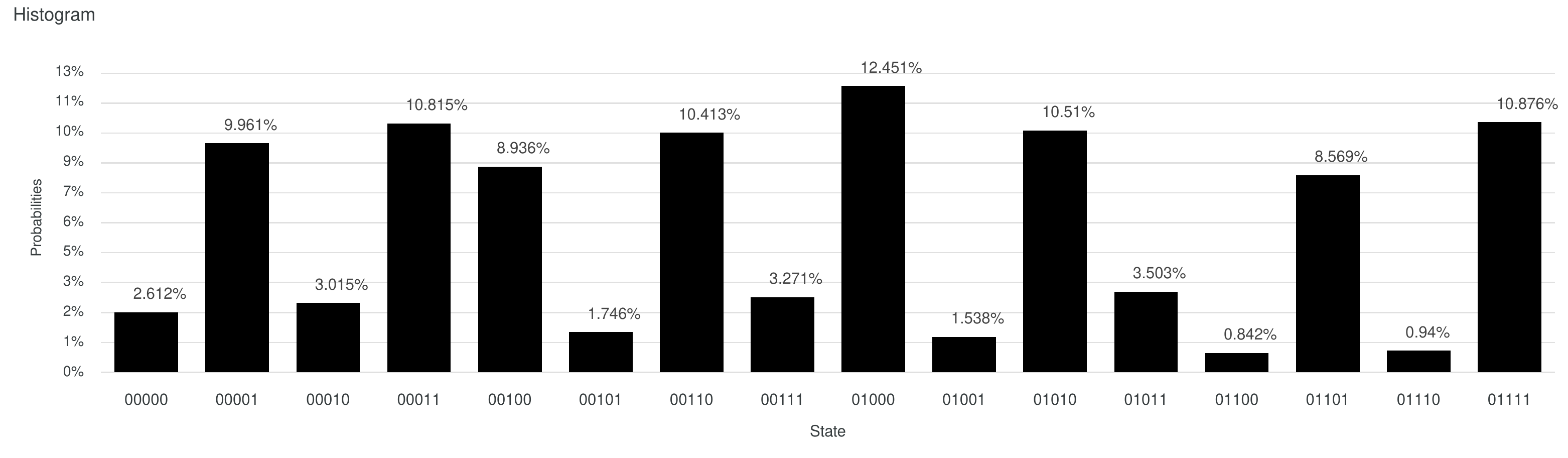}}
 \caption{The probability results of quantum circuit after four measurements in Z-basis. (a) is on $simulator$ and (b) is on $ibmqx2$.}
 \label{ES_Bell_result}
 \end{figure}
Based on the theoretical scheme, we provide the corresponding IBM quantum circuit that can be seen in Figure \ref{ES_Bell_circuit} to generate Bell state by performing three-step quantum walks on $2$-complete graph.
And the results, measured in Z-basis, are shown in Figure \ref{ES_Bell_result}.
According to the Figure \ref{ES_Bell_result_simulator}, the simulation seems very close to the theoretical result (there are eight outcomes appeared in Figure \ref{ES_Bell_result_simulator} and the probability of each is $12.5\%$). And the results on $ibmqx2$ shown in Figure \ref{ES_Bell_result_ibmqx2} are not as good as the simulation results. In fact, there are some noises on the quantum platform, such as decoherence, depolarizing, general noises and so on, which result in the discrepancy between theoretical values and the run results.
According to the quantum state tomography \cite{james2001measurement} and the detailed experimental data, we find that the fidelity between theoretical and simulation density matrix $F(\rho_{ij}^{T},\rho_{ij}^{E\_simulator})$ where $i,j\in\{0,1\}$ are approximately equal to 1 (see Supplementary Note 2).
And the fidelity between theoretical and experimental density matrix $F(\rho_{ij}^{T},\rho_{ij}^{E\_ibmqx2})$ are 0.8535, 0.8793, 0.7909, 0.8549 (see Supplementary Note 2).

\subsubsection{Experimental generation of three-qubit GHZ state on $2$-line}

\begin{figure}[htb]
 \centering
 \subfigure[]{\label{ES_GHZ_circuit}
 \includegraphics[width=7cm]{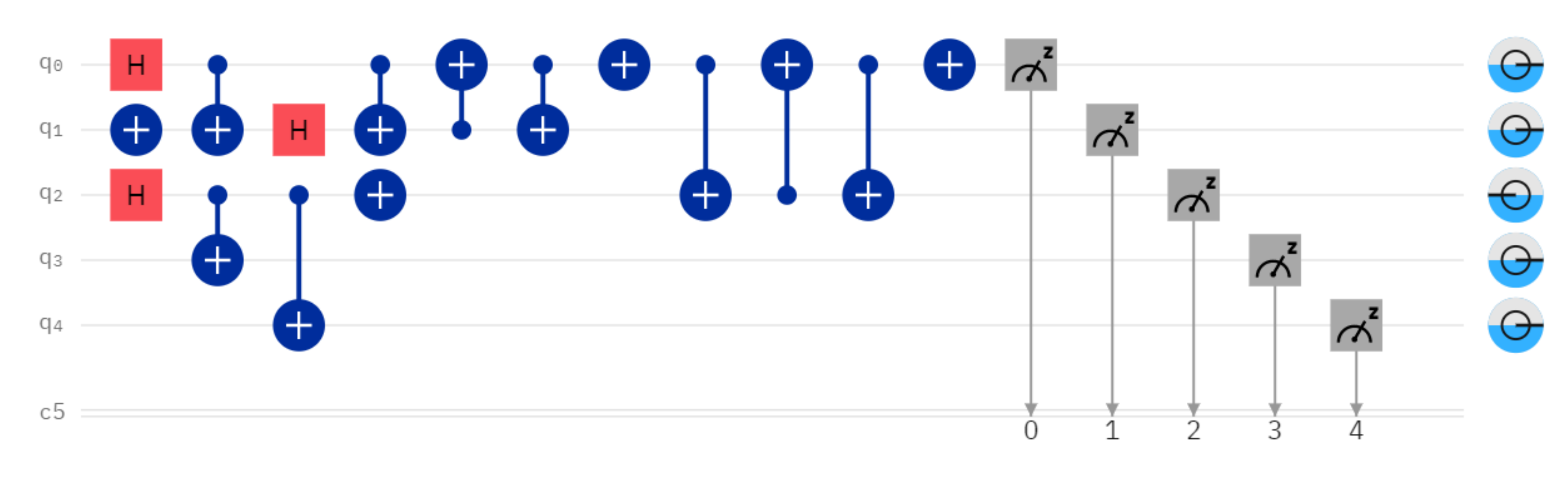}}\ \
 \subfigure[]{\label{ES_GHZ_visualization}
 \includegraphics[width=7cm]{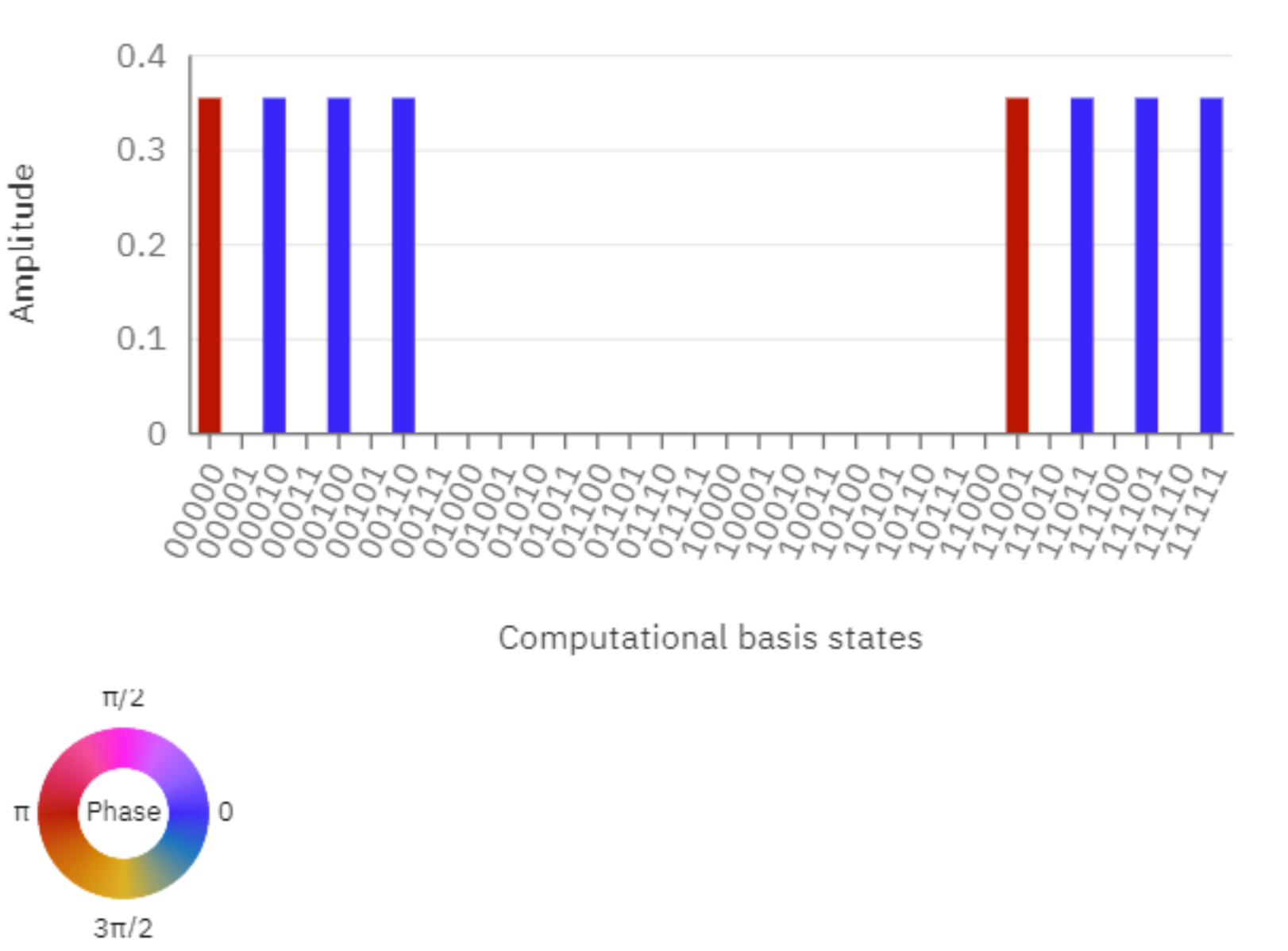}}\ \
 \subfigure[]{\label{ES_GHZ_ibmqx2}
 \includegraphics[width=11cm]{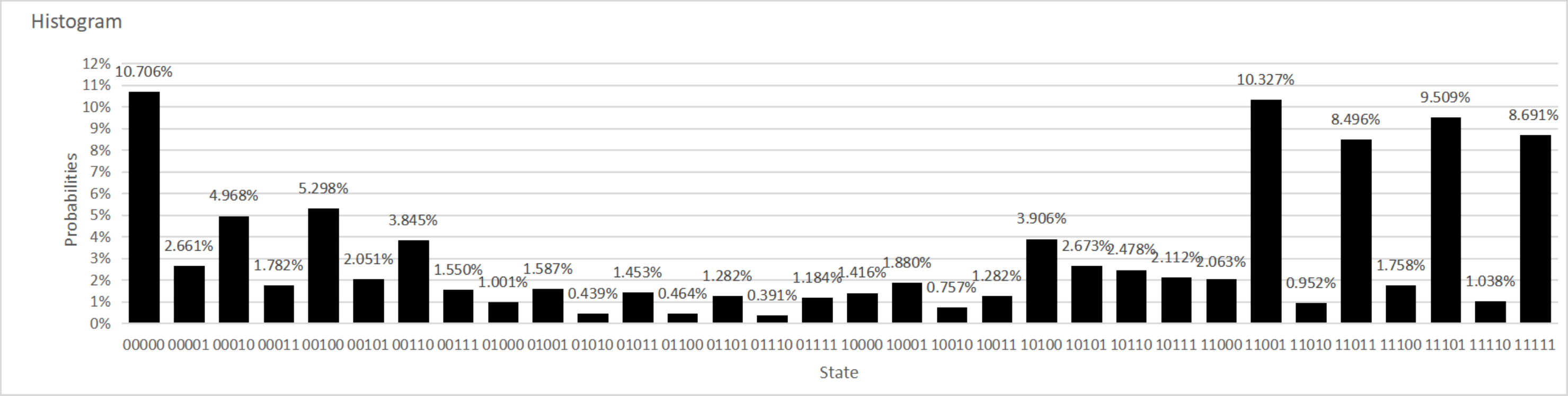}}
 \caption{(a): Quantum circuit for generating three-qubit GHZ state between three parties that are not related at first. (b): The statevector amplitude results of quantum circuit after five measurements in Z-basis.(c): The probability results of quantum circuit on \textit{ibmqx2}.}
 \label{ES_GHZ}
 \end{figure}

As for the generation of three-qubit GHZ state, we provide the corresponding IBM quantum circuit diagram that can be seen in Figure \ref{ES_GHZ_circuit}, which is realized by performing two-step quantum walks on $2$-line on the basis of theoretical scheme.
The whole design idea is similar to the experimental realization of Bell state.
The only thing that needs to be paid attention to is the design of shift operator on $2$-line.
Here we give the amplitude results and probability results on \textit{ibmqx2} in Figure \ref{ES_GHZ_visualization} and \ref{ES_GHZ_ibmqx2} respectively.
The analysis of the experimental results can be similarly obtained, so we will not repeat it.

\section{Application}
In order to share secret among some agents and reconstruct it only when they collaborate together, the scheme of quantum secret sharing (QSS) was first proposed in 1999 \cite{mark1999}.
Zhang and Man proposed a multiparty quantum secret sharing (MQSS) protocol by using entanglement swapping \cite{zhang2005MQSS}.
Then Lin and Zhang revised and improved the protocol respectively \cite{lin2007MQSS,zhang2007reply}.

Here, we improve and perfect the performance of MQSS protocol proposed by Zhang and Man \cite{zhang2005MQSS} by using our scheme.
According to our scheme, for the initial state
\begin{equation}
|\phi(0)\rangle=|\psi_{x,y}\rangle_{1,2}|\psi_{k,l}\rangle_{3,4}=\frac{1}{d}\sum_{m,n=0}^{d-1}\exp(\frac{2\pi \mathrm{i}}{d}mx+nk)|m,m-y,n,n-l\rangle_{1,2,3,4},
\end{equation}
after three-step quantum walks (the coin operators are $I$, $F$, $X_{d}$ in sequence), we could obtain the final state
\begin{equation} \label{application}
\begin{split}
|\phi(3)\rangle&=\frac{1}{d\sqrt{d}}\sum_{m,n,p=0}^{d-1}\exp(\frac{2\pi \mathrm{i}}{d}mx+nk+np)|2m+p+n-y-l+1,m-y,p,n-l+1\rangle_{1,2,3,4} \\
&=\frac{1}{d}\sum_{m,p=0}^{d-1}\exp(\frac{2\pi \mathrm{i}}{d}(m x-(2m+p-y-l+1)(k+p)))|m-y\rangle_{2}|p\rangle_{3}|\psi_{k+p,2m+p-y}\rangle_{1,4}.
\end{split}
\end{equation}
Assume that $d$ is odd, we can set up a bijection between the pairs $(m-y, p)$ and $(k+p, 2m+p-y)$, where $y$ and $k$ are known fixed numbers.

\begin{figure}[htb]
 \centering
 \includegraphics[width=10cm]{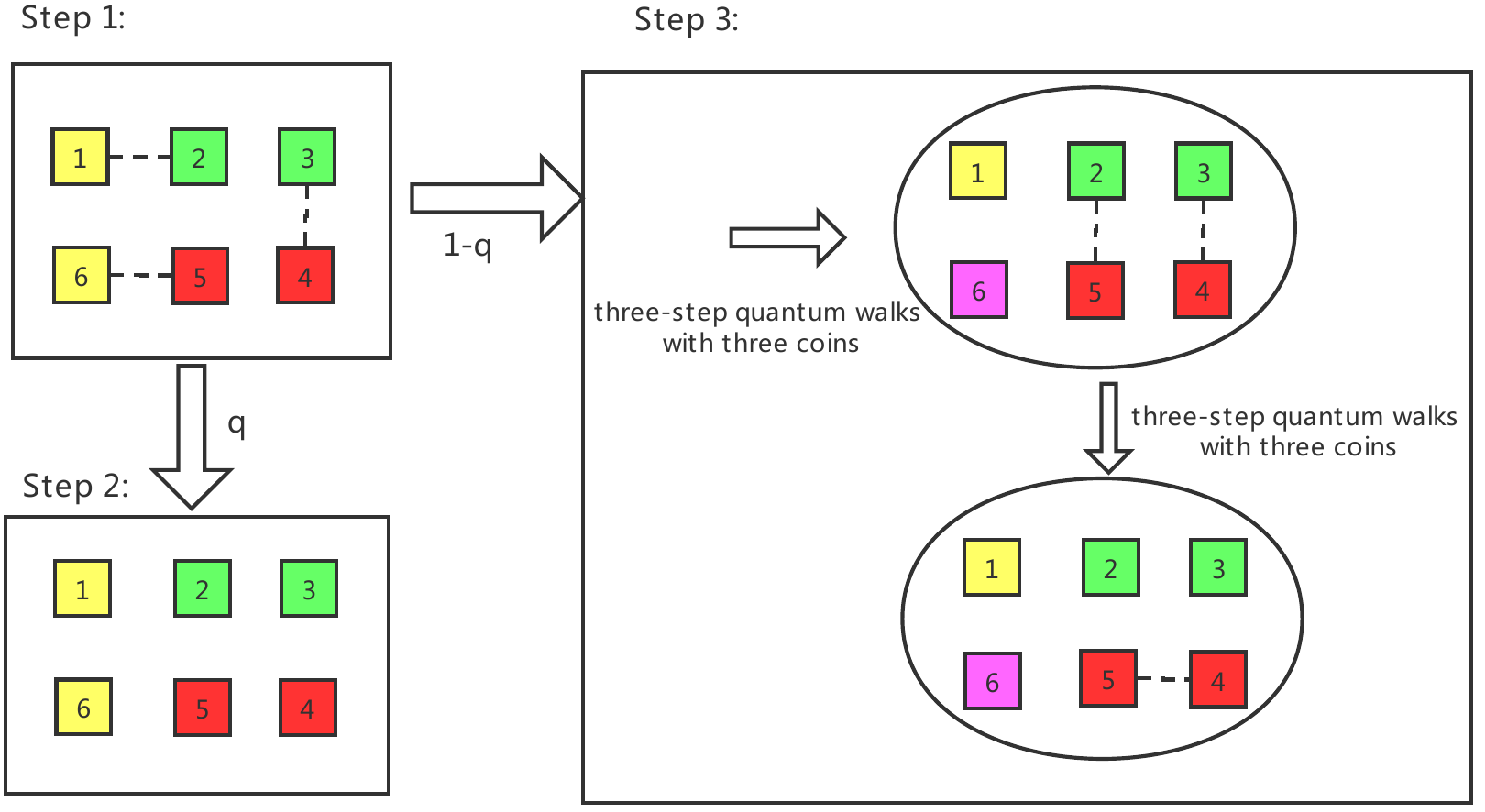}
 \caption{The schematic diagram of the improved multiparty quantum secret sharing where the small boxes indicate particles and the dotted lines indicate entanglement. The particles in yellow belong to Alice. The particles in green and red are in Bob's and Charlie's possession. The purple particle has been performed by a unitary operator.}
 \label{diagram_application}
 \end{figure}

Let $U_{i}=X_{d} ^{i}=\sum_{j=0}^{d-1}|j+i\rangle\langle j|$ be $d$ local unitary transformations ($i=0,1,\cdots,d-1$) that correspond to $\lceil\log d\rceil$ classical bits respectively. To be specific, $U_{0}$ corresponds to $``0\cdots000"$, $U_{1}$ corresponds to $``0\cdots001"$, $U_{2}$ corresponds to $``0\cdots010"$ and so on.
Based on the property $U_{i}$ holds,
\begin{equation}
(I\otimes U_{i})|\psi_{k,l}\rangle=|\psi_{k,l-i}\rangle,
\end{equation}
we can establish correspondence between the unitary operators $\{U_{i}: i=0,1,\cdots,d-1\}$ and generalized Bell states $\{|\psi_{k,l}\rangle: l=0,1,\cdots,d-1\}$ for an arbitrary fixed $k$.
Next, we will show the improved MQSS based on \cite{zhang2005MQSS} in which $d$ is an odd number. And the schematic flow chart of the whole process is shown in Figure \ref{diagram_application}.

\begin{enumerate}
  \item Alice prepares three identical generalized Bell state pairs $|\psi_{k,l}\rangle_{1,2}$, $|\psi_{k,l}\rangle_{3,4}$ and $|\psi_{k,l}\rangle_{5,6}$. She sends qudit 2 and 3 to Bob, 4 and 5 to Charlie via their quantum channels respectively. Then, Alice enters step 2 with probability $q$, or enters step 3 with probability $1-q$.
  \item Alice chooses randomly from two mutually unbiased measurement basis, $\mathcal{M}_{1}=\{|i\rangle: i=0,1,\cdots,d-1\}$ and $\mathcal{M}_{2}=\{|\tilde{i}\rangle: i=0,1,\cdots,d-1\}$, where $|\tilde{i}\rangle=\frac{1}{\sqrt{d}}\sum_{j=0}^{d-1}\exp(\frac{2\pi \mathrm{i}}{d}ij)|j\rangle$. Alice tells Bob the basis she has chosen. They measure qudit 1 and 2 using this chosen measurement basis respectively. Then Bob tells Alice his outcome of qudit 2. Alice compares these two outcomes to judge the security of Alice-Bob quantum channel (For example, if the generalized Bell state pairs that Alice prepares is $|\psi_{0,0}\rangle$, these two measurement outcomes must be identical when the channel is safe). The security of Alice-Charlie quantum channel can also be checked in the same way. If the two channel are both secure, returns to step 1. Otherwise, this process will be terminated.
  \item Alice applies a local unitary operator chosen randomly from $\{U_{i}: i=0,1,\cdots,d-1\}$ on one of her qudits 1 and 6 ( say, on qudit 6). Firstly, they perform three-step quantum walks on qudit 5, 6, 1 and 2. Alice performs two local measurements on qudit 6 and 1 and announces her outcome. Next, they perform three-step quantum walks on qudit 5, 2, 3 and 4. Bob performs two local measurements on qudit 2 and 3 and writes down the outcomes. At this times, Charlie records his own state. Bob and Charlie can deduce the unitary transformation Alice performs and the classical bits Alice wants to share.
\end{enumerate}

Now let us explain how we derived $U_{i}$ in step 3 based on the three measurement results they announced and evolutionary properties.
Without generality, we can denote the outcome Alice announces as $|a,b\rangle_{6,1}$.
Suppose the quantum state of qudit 5 and 2 is $|\psi_{\bar{k}, \bar{l}}\rangle$ after the first quantum walks.
According to equation (\ref{application}), we can denote the outcome Bob writes down and the state Charlie possesses as $|m-\bar{l},p\rangle_{2,3}$ and $|\psi_{k+p,2m+p-\bar{l}}\rangle_{5,4}$ respectively.
Since $k$ is known, $p$, $m$ and $\bar{l}$ are also known.
Next we can infer that $\bar{k}=k+p$ and the state of qudit 5 and 6 after the local unitary transformation performed by Alice is $|\psi_{k,\bar{l}-b-2a}\rangle_{5,6}$ based on the first quantum walks.
Thus, the unitary transformation Alice performed is $U_{l-(\bar{l}-b-2a)}$. Note that the above mentioned addition and subtraction methods are all of modulo $d$. So it is not hard to infer the classical bits Alice wants to share.

In addition, our protocol can also be generalized to a multiparty case. Without loss of generality, we suppose that there are $N$ parties in total.
First, the sender prepares $N$ identical generalized Bell state pairs and distributes two qudits to the agents one by one and does not tell their the ordering of the two qudits. So the $i$-th agent and the $i+1$-th agent share a generalized Bell state pair. And the $N$-th agent will share a generalized Bell state pair with the sender.
Then, they check the security of quantum channel (step 2) with probability $q$ or transmit message (step 3) with probability $1-q$. The order of quantum walk and measurements, which is crucial for the whole process, is the same as the order of entanglement swapping in the previous protocol \cite{lin2007MQSS, zhang2007reply}.
The security (even in a noisy channel) can be guaranteed by the discussions in Ref. \cite{zhang2005MQSS, lin2007MQSS, zhang2007reply}.

Now, let us make some comparisons between our scheme and the other schemes \cite{zhang2005MQSS, zhang2007qutrit}.
Compared with Ref. \cite{zhang2005MQSS}, the advantage of our scheme is that the capacity of classical messages it can distribute is much larger.
Specifically, it is a task to distribute $N=\lceil\log d\rceil$ classical bits instead of just two bits.
In addition, Ref. \cite{zhang2005MQSS} needs $\frac{3N}{2(1-q)}$ Bell states in step 3. Our scheme just needs $\frac{3}{1-q}$ generalized Bell states. Thus, as for the resource, our scheme is much better.
In Ref. \cite{zhang2007qutrit}, Zhang et.al designed a MQSS protocol by swapping qutrit-state entanglement and then generalized to the qudit case, in which the so-called two-qutrit (two-qudit) entangled-state measurement is crucial but hard to implement in experiment.
Here, we mainly use the model of quantum walks with multiple coins which is practical and novel.
Therefore, our scheme avoids the difficulty in the experimental realization of Bell state measurement, especially for the measurement of high dimensional Bell state \cite{John2002generalizedmeasurment}.
So our protocol is more feasible than the protocol in Ref. \cite{zhang2007qutrit}.
Also, it may inspire more potential applications in quantum communication and quantum information.

\section{Discussion}
The generation of entanglement, especially in the high-dimensional case, between several designated parties is vital and essential in scalable quantum network.
Inspired by entanglement swapping, we have developed a scheme to generate entangled state by using the model of quantum walks with multiple coins, including two-qubit entangled state, two-qudit entangled state, three-qubit GHZ state and three-qudit GHZ state.
In comparison with original entanglement swapping, the benefit of using our scheme is expected to avoid performing joint Bell state measurement that is still an unsolved problem to date, and provide a relatively general framework for generating entangled state via entanglement swapping.

Also, we provide the experimental realization of our scheme to generate Bell state and three-qubit GHZ state on IBM platform.
And our scheme exhibits high fidelity by performing quantum state tomography.
Moreover, we present an improved multiparty quantum secret sharing by using our our scheme.
After our improvement, it can convey more classical information and require less quantum resources (Bell states).

It remains an interesting open question to design further the GHZ state and $W$ state for a system of $n$ qubit, where $n$ is an arbitrary integer.
It is also highly desired to extend our method to more practical applications.

\section{Data availability}
The experimental data that supports our finding are available from the corresponding author on reasonable request.

\section{Acknowledgements}
We thank the support of National Key Research and Development Program of China under grant 2016YFB1000902, National Natural Science Foundation of China (Grant No.61472412, 61872352), and Program for Creative Research Group of National Natural Science Foundation of China (Grant No. 61621003).

\end{document}